# Factors correlating to enhanced surface diffusion in metallic glasses

Ajay Annamareddy,[1,a)] Yuhui Li,[2] Lian Yu,[2,3] Paul M. Voyles,[1] and Dane Morgan[1,a)]

[1]Department of Materials Science and Engineering, [2]School of Pharmacy and [3]Department of Chemistry, University of Wisconsin-Madison, Madison, WI 53706, USA.

a) Authors to whom correspondence should be addressed: vkannama@ncsu.edu and ddmorgan@wisc.edu

The enhancement of surface diffusion ($D_S$) over the bulk ($D_V$) in metallic glasses (MGs) is well documented and likely to strongly influence the properties of glasses grown by vapor deposition. Here, we use classical molecular dynamics simulations to identify different factors influencing the enhancement of surface diffusion in MGs. MGs have a simple atomic structure and belong to the category of moderately fragile glasses that undergo pronounced slowdown of bulk dynamics with cooling close to the glass transition temperature ($T_g$). We observe that $D_S$ exhibits a much more moderate slowdown compared to $D_V$ when approaching $T_g$, and $D_S/D_V$ at $T_g$ varies by two orders of magnitude among the MGs investigated. We demonstrate that both the surface energy and the fraction of missing bonds for surface atoms show good correlation to $D_S/D_V$, implying that the loss of nearest neighbors at the surface directly translates into higher mobility, unlike the behavior of network- and hydrogen-bonded organic glasses. Fragility, a measure of the slowdown of bulk dynamics close to $T_g$, also correlates to $D_S/D_V$, with more fragile systems having larger surface enhancement of mobility. The deviations observed in the fragility – $D_S/D_V$ relationship are shown to be correlated to the extent of segregation or depletion of the mobile element at the surface. Finally, we explore the relationship between the diffusion pre-exponential factor ($D_0$) and activation energy ($Q$) and compare to a $\ln(D_0)$-$Q$ correlation previously established for bulk glasses, demonstrating similar correlations from MD as in the experiments and that the surface and bulk have very similar $\ln(D_0)$-$Q$ correlations.

## I. Introduction

The enhancement of in-plane surface diffusion coefficient ($D_S$) over the bulk (or volume) diffusion coefficient ($D_V$) in metallic glasses (MGs) is a property of fundamental interest with potential application to engineering so-called ultrastable glasses (UGs) by physical vapor deposition. By carefully optimizing the substrate temperature and deposition rate, Ediger *et al.* have created organic UGs with exceptional thermodynamic and kinetic stability.[1] Glasses obtained using a traditional melt-quench approach need to be aged for an impractical amount of time to obtain similar stability.[2] The metallic counterpart of UGs was first created by Yu *et al.* and found to have excellent properties like high elastic modulus.[3] UGs form under deposition due to the additional relaxation enabled by the larger value of $D_S$ vs. $D_V$, which allows the as-deposited atoms to have sufficient mobility to quickly find stable, low-energy configurations.[4] Therefore, enhanced surface mobility is necessary to the formation of UGs, and better control of $D_S/D_V$ could provide a pathway to the formation of improved UGs.

Early studies on the measurement of $D_S$ in glasses employed the surface grating decay method, in which periodic gratings are etched into surfaces and the rate of flattening of the gratings is analyzed. According to the Mullins model,[5] different near-surface mass transport mechanisms can be identified based on their unique relation between decay constant and grating periodicity. In these studies, the surface diffusion enhancement factor, $D_S/D_V$, is as high as $10^8$ at the bulk glass transition temperature $T_g$.[6–10] Enhanced surface mobility is generally attributed to the smaller number of bonds a surface atom breaks when moving compared to a bulk atom. Network glass-formers like silica, which requires the breaking of network oxide bonds to flow, do not show fast surface diffusion during grating decay.[11] Simulations have found that the Si-O bonds are largely intact at the surface as the radial distribution of oxygen atoms around silicon from the bulk to the surface is almost unchanged.[12] Organic materials with extensive hydrogen-bonding also have low surface enhanced diffusion, which is attributed to the preservation of hydrogen bonds close to the surface.[7] In both these cases, there is no significant change of local environment between the bulk and the surface, and the materials do not show a large decrease of the diffusion barrier for the surface atoms.

In this work, we calculate $D_S/D_V$ in metallic glasses (MGs) using classical molecular dynamics simulations. Our main goal is to identify different properties that correlate to $D_S/D_V$ across several MGs and to aid in the selection of materials for designing UGs in the future. Compared to network glasses, MGs significantly disrupt their basic nearest-neighbor bonding at the surface, suggesting that a loss of neighbors could directly translate into an increase of mobility. With this in mind, we first explored how the loss of first neighbors and changes in surface energy correlate to surface enhanced diffusion. As a metallic liquid is supercooled, $D_V$ deviates from $D_S$ and undergoes a marked slowdown while approaching $T_g$, quantified by the fragility index ($m$). In contrast, $D_S$ undergoes a more moderate slowdown towards $T_g$. Based on this, we correlate $D_S/D_V$ to $m$,[7] and find generally good trends but that the correlation breaks down when surface diffusion is significantly impacted by surface segregation. Finally, we use extrapolation methods to approximate the surface enhanced diffusion expected for typical experimentally cooled glasses.

## II. Simulation Methods

Classical molecular dynamics (MD) simulations are employed to study various model MGs in this work. The typical MD simulation timescale is usually less than 1 µs and very high cooling rates compared to experiments need to be employed when preparing glasses in MD. As a result, the glasses created by MD simulations are highly under-relaxed relative to real glasses, and the MD glass dynamics are many orders of magnitude faster than in the real glass. The simulations also cover only 4–5 decades of slowdown in dynamics from the high temperature liquid. Our measured $D_S/D_V$ values are therefore much smaller than typical experimental observations. Nonetheless, using MD simulations, it is now possible to access long enough timescales that many features associated with glass transition physics can be observed, including VFT behavior in the supercooled liquid and Arrhenius behavior in the glass. Therefore, consistent with many researchers in glass physics,[13,14] we will assume that the MD relaxed glass yields mechanisms of bulk and surface diffusion similar to those found in real glasses.

In this work, we study 10 binary MGs that are all modeled by embedded atom method (EAM) potentials to describe the interatomic interactions. The compositions, in terms of mole percent, include $Cu_XZr_{100-X}$ (X= 25, 35, 50, 65, 75, and 90),[15] $Pd_{80}Si_{20}$,[16] $Ni_{80}P_{20}$,[16] $Al_{90}Sm_{10}$,[17] and $Ni_{33}Zr_{67}$.[18] To generate a glass, the samples were initially equilibrated for 2 ns at 2000 K. The resulting liquids were then cooled to 1000 K in 50 K decrements at the rate of 100 K per 6 ns. During this phase, the systems were sufficiently equilibrated, as the structural relaxation times are of the order of ps or less. Subsequently, the samples were cooled to 500 K in 20 K decrements at the rate of 100 K per 60 ns, corresponding to a cooling rate of ~$10^9$ K/s. All systems undergo the glass transition during this stage, at the temperatures listed in Table 1, as identified from the change in slope observed in the temperature variation of enthalpy or volume. Compared to the experimental $T_g$, also listed in Table 1 for the compositions we have found in the literature, MD glass transition temperatures are usually higher due to orders of magnitude faster cooling during the simulation, although some errors from the EAM potentials are also expected. NPT conditions with zero nominal pressure were employed during the whole procedure. A cubic simulation box with 16384 atoms was used and periodic boundary conditions (PBC) were applied in all three Cartesian directions. The time step applied was 1 fs. $Cu_{10}Zr_{90}$ was also tried but found to crystallize with the above protocol and is not part of this study.

To study the bulk diffusion, the final configuration at the temperature of interest during the quenching process was used as the starting point for production runs in NVT conditions with PBC in all directions. The correct volume for the NVT simulation was obtained by using the volume from the NPT quenching. To measure surface diffusion, with the same configuration obtained from quenching process as before, free surfaces were created by extending the simulation cell boundaries by 10 Å along the $\pm$z-axis. Again, NVT conditions and PBC (although the atoms cannot interact through the boundaries in the z-direction) were applied and the system was initially equilibrated for 1 ns to make sure the newly created surfaces at the two edges were relaxed before the production phase begins. Atoms in the outer 2.5 Å (or 12.5 Å of the cell considering the empty space), which is approximately the thickness of a monolayer as determined from radial distribution function of atoms, along the $\pm$z-axis were used to evaluate the properties associated with the surface. While the outer monolayer at the free boundary has the highest atomic mobility and is chosen as representing the surface in this work, there is an interfacial region of perhaps up to a few nm thickness exhibiting gradients in mobility from surface towards the bulk. The thickness of the interfacial region varies with temperature, and MD simulations in both glass-forming and crystalline materials have investigated this behavior extensively using metrics such as Debye-Waller factor.[19–21] In $Cu_{64}Zr_{36}$ glass films, a non-monotonic variation of interface thickness with temperature has been observed.[19]

The diffusion coefficient, $D$, is calculated based on Einstein's equation that relates $D$ to the mean-squared displacement (MSD) of atoms as

$$D = \lim_{t \to \infty} \frac{1}{(2d)Nt} \left\langle \sum_{i=1}^{N} |r_i(t) - r_i(0)|^2 \right\rangle, \quad (1)$$

where $d$ is the number of dimensions, $N$ is the number of atoms, $r_i(t)$ is the position of atom $i$ at time $t$, and the angular brackets refer to the ensemble average. Fig. S1 in the Supplementary Material shows the time evolution of MSD in bulk $Cu_{50}Zr_{50}$ system at two temperatures in the glassy state. For the bulk diffusion coefficient, $d = 3$. For surface diffusion, only the lateral displacement of atoms in the xy-plane is considered with $d = 2$, and atoms belonging to the surface layer at both $t = 0$ and $t = t$ are considered as contributing to the summation in Eq. (1). Unless explicitly noted, the diffusion values in this study are averaged over all the atoms and, hence, are composition-averaged values. Relaxation time, $\tau$, of the bulk is extracted from self-intermediate scattering function, defined as

$$F_s(q,t) = (1/N) \left\langle \sum_{n=1}^{N} e^{i\mathbf{q} \cdot [\mathbf{r}_n(t) - \mathbf{r}_n(0)]} \right\rangle,$$

and $F_s(q,\tau) = 1/e$. The magnitude of the chosen wave vector $q$ coincides with the first maximum of the static structure factor. Fig. S2 in the Supplementary Material shows an illustration of the decay of $F_s$ with time. The lowest temperature studied in the glassy state, where $\tau$ grows very fast with cooling, is chosen such that an MSD of at least 5 Å$^2$ per atom is obtained during a simulation time of a few 100s of ns. We also compare the lowest temperature simulation times to bulk relaxation time for all the MGs in Table S1 in the Supplementary Material. For equilibrium liquids, the Stokes-Einstein (SE) equation relates $D$ to $\tau$ by $D/T \sim (1/\tau)$, where $T$ is the temperature. In the supercooled liquid and glassy states, the SE equation breaks down indicating the decoupling of diffusion and relaxation time, and the relation between $D$ and $\tau$ is expressed in terms of the fractional SE equation represented as $D/T \sim (1/\tau)^{1-\zeta}$ where $\zeta$ is the decoupling parameter.[22] In Fig. S3 of the Supplementary Material, we show a plot of $D/T$ versus $1/\tau$ that is used to estimate $\zeta$, and, in Table 1, we give the value of $\zeta$ for all the MGs studied in this work. To improve statistics, the method of buffer averaging[23] is employed for both diffusion and relaxation time calculations.

TABLE 1. Calculated glass transition temperature ($T_g$), surface enhanced diffusion ($D_S/D_V$) at $T_g$, and decoupling parameter ($\zeta$) using MD simulations for all the MGs studied. Experimental $T_g$ are shown for the compositions we have found in the literature.

| Composition ($A_XB_{100-X}$) | MD $T_g$ (K) | Expt. $T_g$ (K) | $D_S/D_V$ at $T_g$ | Decoupling parameter ($\zeta$) |
|---|---|---|---|---|
| $Cu_{25}Zr_{75}$ | 720 | - | 63.06 | 0.29 |
| $Cu_{35}Zr_{65}$ | 700 | 616[24] | 37.16 | 0.27 |
| $Cu_{50}Zr_{50}$ | 700 | 666[24] | 26.09 | 0.28 |
| $Cu_{65}Zr_{35}$ | 760 | 737[24] | 28.57 | 0.36 |
| $Cu_{75}Zr_{25}$ | 780 | - | 25.86 | 0.40 |
| $Cu_{90}Zr_{10}$ | 740 | - | 10.54 | 0.41 |
| $Pd_{80}Si_{20}$ | 760 | 657[25] | 117.54 | 0.33 |
| $Ni_{80}P_{20}$ | 560 | 580[25] | 1.40 | 0.37 |
| $Al_{90}Sm_{10}$ | 640 | 445[26] | 14.29 | 0.48 |
| $Ni_{33}Zr_{67}$ | 660 | - | 11.91 | 0.27 |

## III. Results

### 1. Surface enhanced diffusion in different metallic glasses

We begin with describing the typical evolution of dynamics with cooling in MGs. Fig. 1(a) shows the temperature variation of surface and bulk diffusion in $Cu_{50}Zr_{50}$ in an Arrhenius plot. In the high-temperature liquid state, $D_S$ and $D_V$ are comparable. As the alloy is supercooled, a significant slowdown in the bulk dynamics is observed when approaching the glass transition temperature $T_g$ that can be described by the well-known Vogel-Fulcher-Tammann (VFT) equation.[27] In contrast, the surface undergoes only a modest slowdown. Below $T_g$, identified as 700 K from the temperature dependence of volume or enthalpy, both $D_S$ and $D_V$ can be fitted by the Arrhenius equation, $D = D_0 e^{-Q/(k_B T)}$. Here, $D_0$ is the pre-exponential factor and $Q$ is the activation energy. The bulk dynamics of MGs showing VFT behavior in the supercooled liquid regime and Arrhenius behavior below $T_g$ is well known from experiments,[28] and we have observed similar behavior for all the MGs studied in this work. Table 2 shows the values of activation energy and pre-exponential factor in the glassy state for all the MGs in this work. Our simulations below $T_g$ are likely far from the most stable glassy state and it is not clear a priori that their diffusion should be Arrhenius. However, if we assume that the structure of the glass undergoes minimal changes with temperature and also during the time scales of the MD simulations then it is reasonable to expect that the systems see an almost fixed distribution of activation energies and will show approximately Arrhenius behavior. The assumption that structure is fairly constant during the MD timescales is supported by the fact that we obtain linear MSD with time for all our MD simulations and that timescales for significant structural relaxation in glasses from experiments are typically minutes, hours, or even years. To further support our claim, we reran the $D$ value for the highest temperature in the glassy state ($T = 700K$) in $Cu_{50}Zr_{50}$ starting from the final atomic configuration obtained after the first determination of $D$, which represented 30 ns of ageing. We obtained a value just ~5 % lower from that obtained from our first determination of $D$ at 700K. Our Arrhenius behavior is also consistent with other MD simulations.[29] $D_S/D_V$ at $T_g$ in the case of $Cu_{50}Zr_{50}$ is ~26. Across the 10 different MGs we have studied, $D_S/D_V$ varies by almost two orders of magnitude, as listed in Table 1 and shown in Fig. 1(b).

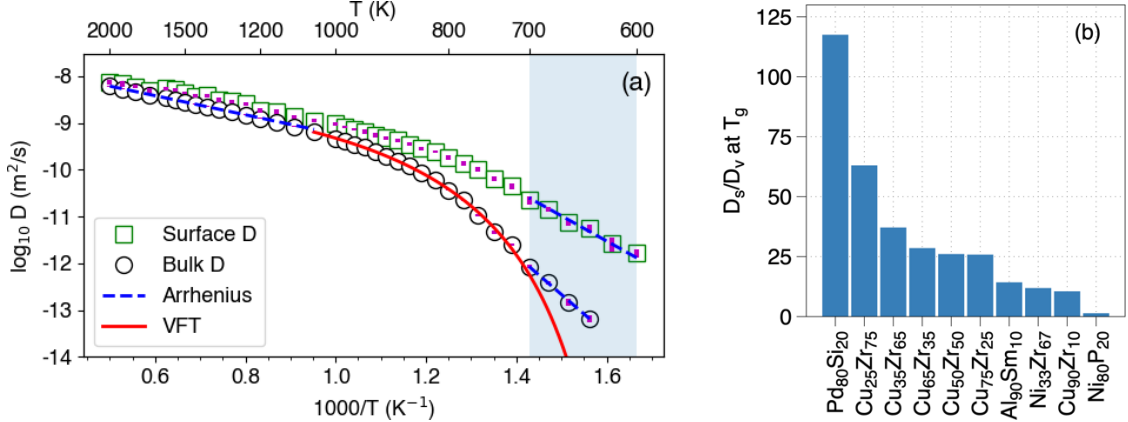

FIG 1. (a) Plot of surface and bulk diffusion as a function of temperature for $Cu_{50}Zr_{50}$. The pronounced slowdown of bulk dynamics in the supercooled regime is well described by the VFT equation. Below $T_g$, both surface and bulk diffusion satisfy the Arrhenius equation. The standard deviations in $D$ (from 5 independent simulations) are shown in pink and they are smaller than the symbol size at all the temperatures studied. (b) Surface-enhanced diffusion ($D_S/D_V$) of different MGs investigated in this study. The compositions along the x-axis, in the order of decreasing $D_S/D_V$, are $Pd_{80}Si_{20}$, $Cu_{25}Zr_{75}$, $Cu_{35}Zr_{65}$, $Cu_{65}Zr_{35}$, $Cu_{50}Zr_{50}$, $Cu_{75}Zr_{25}$, $Al_{90}Sm_{10}$, $Ni_{33}Zr_{67}$, $Cu_{90}Zr_{10}$, and $Ni_{80}P_{20}$.

TABLE 2. MD calculated bulk glass activation energy ($Q_V$), surface glass activation energy ($Q_S$), bulk pre-exponential factor ($D_{0,V}$), and surface pre-exponential factor ($D_{0,S}$) for diffusion in different binary ($A_xB_{100-x}$) glasses. For each composition, the three rows of values correspond to total, component $A$, and component $B$ diffusion.

| Composition ($A_xB_{100-x}$) | $Q_V$ (eV) | $Q_S$ (eV) | $D_{0,V}$ (m$^2$/s) | $D_{0,S}$ (m$^2$/s) |
|---|---|---|---|---|
| $Cu_{25}Zr_{75}$ (total) | 1.47 | 0.96 | 0.011 | 8.2e-5 |
| component $A$: | 1.42 | 0.93 | 0.010 | 9.0e-5 |
| component $B$: | 1.53 | 1.00 | 0.022 | 1.0e-4 |
| $Cu_{35}Zr_{65}$ | 1.53 | 1.06 | 0.804 | 0.072 |
| | 1.47 | 0.99 | 0.355 | 0.022 |
| | 1.66 | 1.12 | 6.189 | 0.189 |
| $Cu_{50}Zr_{50}$ | 1.66 | 0.93 | 1.486 | 2.0e-4 |
| | 1.61 | 0.91 | 0.837 | 1.3e-4 |
| | 1.86 | 1.07 | 24.199 | 7.5e-4 |
| $Cu_{65}Zr_{35}$ | 1.36 | 0.84 | 0.001 | 1.9e-5 |
| | 1.33 | 0.81 | 0.001 | 1.4e-5 |
| | 1.57 | 0.99 | 0.021 | 6.1e-5 |
| $Cu_{75}Zr_{25}$ | 1.40 | 0.76 | 0.005 | 5.4e-6 |
|  | 1.38 | 0.75 | 0.004 | 4.5e-6 |
|  | 1.73 | 0.88 | 0.340 | 4.3e-6 |
| $Cu_{90}Zr_{10}$ | 1.23 | 0.91 | 0.001 | 1.4e-4 |
|  | 1.23 | 0.90 | 0.001 | 1.2e-4 |
|  | 1.61 | 1.24 | 0.099 | 0.014 |
| $Pd_{80}Si_{20}$ | 1.43 | 1.10 | 0.010 | 0.001 |
|  | 1.42 | 1.09 | 0.008 | 0.001 |
|  | 1.74 | 1.75 | 8.103 | 1.571 |
| $Ni_{80}P_{20}$ | 1.11 | 0.70 | 0.052 | 2.6e-5 |
|  | 1.10 | 0.73 | 0.051 | 1.6e-5 |
|  | 1.66 | 0.79 | 2889.512 | 0.008 |
| $Al_{90}Sm_{10}$ | 0.74 | 0.60 | 8.1e-6 | 1.1e-5 |
|  | 0.74 | 0.58 | 1.5e-5 | 8.5e-6 |
|  | 1.09 | 0.85 | 7.1e-4 | 0.033 |

## 2. Diffusion activation energies and pre-exponential factors

As both surface and bulk diffusion follow Arrhenius behavior in the glassy state, the surface enhanced diffusion can be written as

$$\frac{D_S}{D_V} = \frac{D_{0,S} e^{-Q_S/(k_B T)}}{D_{0,V} e^{-Q_V/(k_B T)}} = \frac{D_{0,S}}{D_{0,V}} e^{(Q_V - Q_S)/(k_B T)}. \qquad (2)$$

From Eq. (2), $D_S/D_V$ at $T_g$ depends on the difference in activation energies normalized by the glass transition temperature, $(Q_V - Q_S)/k_B T_g$, and the ratio of pre-exponential factors, $D_{0,S}/D_{0,V}$. If either of these factors is invariant across the MGs, the other can be easily identified as having a direct correlation to $D_S/D_V$. Fig. 2(a) shows the variation of $Q_V$, $Q_S$, and $(Q_V - Q_S)/k_B T_g$, with the MGs on the abscissa arranged such that $D_S/D_V$ decreases from left to right, similar to Fig. 1(b). As expected, $Q_V$ is always greater than $Q_S$, although in some cases they are comparable to each other. A monotonic trend in $(Q_V - Q_S)/k_B T_g$ is not evident in Fig. 2(a) along the abscissa, indicating that the difference in activation energies does not correlate directly to the measured surface enhanced diffusion, which can be attributed to the variability observed in the ratio of pre-exponentials. In Fig. 2(b), we depict $D_{0,V}/D_{0,S}$ for different MGs, which vary by four orders of magnitude. The variation of the ratio of pre-exponentials is also not monotonic from left to right, and, hence, $D_S/D_V$ is not directly correlated to the ratio of pre-exponentials.

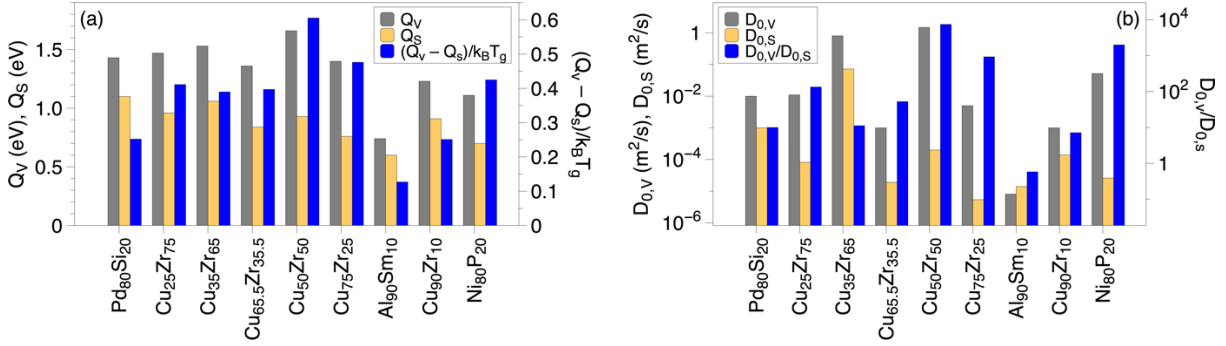

FIG. 2. Correlating surface enhanced diffusion to (a) the normalized difference in activation energies $(Q_V - Q_S)/k_B T_g$ (divided by 20), and (b) the ratio of pre-exponential factors $D_{0,V}/D_{0,S}$. The similar variation of $(Q_V - Q_S)$ and $\log(D_{0,V}/D_{0,S})$ reflects the underlying connection between $Q$ and $D_0$ observed earlier.[30]

$(Q_V - Q_S)$ (ignoring the normalization factor $k_B T_g$ as it only has a small impact) in Fig. 2(a) and $\ln(D_{0,V}/D_{0,S})$ in Fig. 2(b) have similar variation across the MGs studied, which is consistent with the relation between $Q$ and $D_0$ previously established in MGs from experiments. Sharma and Faupel[30] have compiled the data of $Q$ and $D_0$ from several diffusion studies in MGs and indicated a linear relation between $Q$ and $\ln(D_0)$. Fig. 3 replots $Q$ versus $D_0$ taken from Sharma and Faupel, together with the linear fit observed in their study[30] and adds the results from our simulations. Remarkably, both our MD bulk and surface diffusion data show similar slope as the experimental data. However, for the same $Q$, the simulated $D_0$ is many orders of magnitude higher than the experiments, which reflects the underrelaxed nature of MD glasses compared to real glasses and that the MD diffusion coefficient values are many orders of magnitude higher than the experimental diffusion values. We make an approximate correction to remove this effect in Sec. S4 of the the Supplementary Material. We note that the here observed linear correlations between the logarithm of pre-factor and $Q$ were earlier seen in the studies of interfacial regions of crystals[31] and biological preservation of materials.[32]

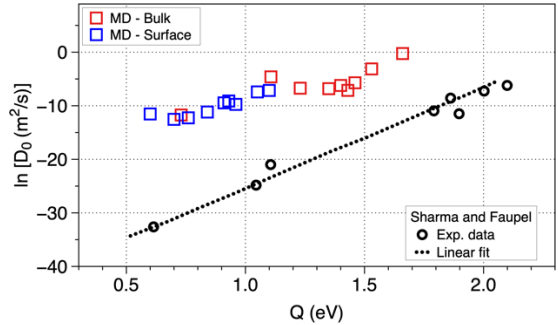

FIG. 3. Plot of activation energy ($Q$) versus pre-exponential factor ($D_0$) for MGs from experiments and simulations. The open circles and the dotted line representing the linear relation between $Q$ and $\ln(D_0)$ are taken from Sharma and Faupel[30] compiling several experimental bulk diffusion studies. Open squares correspond to our MD simulations, with Arrhenius fitting of bulk and surface diffusion below $T_g$ as shown in Fig. 1(a).

### 3. Metrics related to missing bonds

Free surfaces lead to missing bonds for many atoms, with a concomitant increase in energy of the system. This energy increase can be quantified by the surface energy, defined as the difference in energy of the system with free surfaces and energy of the bulk system and normalized by the total area of free surfaces. Given the strong connection of missing bonds to both surface energy and surface diffusion, it is reasonable to consider that they might be correlated. Such a correlation could be useful for providing intuitions about surface diffusion and methods to accelerate the measurement or prediction of surface diffusion. Fig. 4(a) shows how $D_S/D_V$ varies with surface energy for the MGs in our study. As expected, we see that higher surface energy correlates with higher $D_S/D_V$, although there is a significant scatter in the correlation.

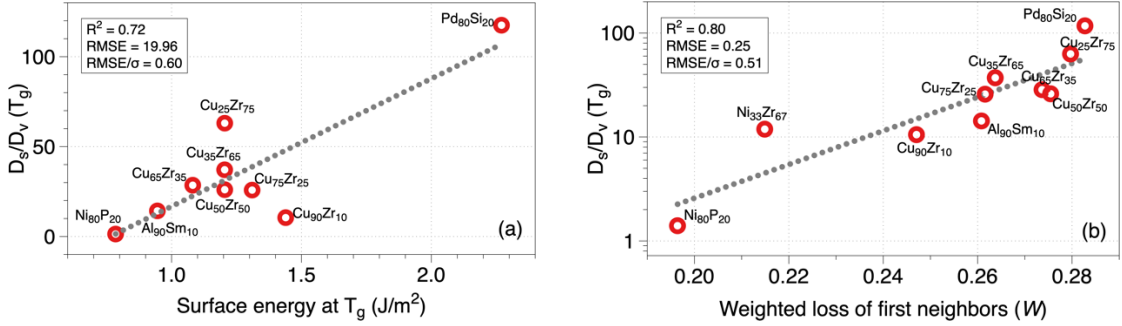

FIG. 4. Correlation of surface enhanced diffusion with (a) surface energy and (b) weighted loss of first neighbors across several MGs in our study. In this work, the accuracy of the correlations is quantified by using standard metrics like $R^2$, root-mean-squared error (RMSE), and RMSE normalized by the standard deviation (RMSE/$\sigma$).

The idea for the smaller number of bonds (or first-neighbors) leading to enhanced interfacial mobility follows from the fact that hopping rates depend on activation energy (this is a result of the Eyring-Polanyi relation or transition state theory) and that activation energies typically scale with on-lattice stability (known as the Bronsted-Evans-Polanyi relation), and on-lattice energies scale with the number of bonds.[33] Here, we attempt to directly measure the missing bonds at the surface and correlate that with $D_S/D_V$. For an alloy it is reasonable to expect that the missing bonds for each species might play a different role, depending on the amount of a species at the surface and its diffusion coefficient. Specifically, one might expect that the importance of a missing bond would scale with the diffusion coefficient of the atom type with the missing bond. For example, if the surface primarily breaks the bonds of the slower species, then it might have a much smaller impact on $D_S/D_V$ than if it primarily breaks the bonds of the faster species. With this motivation, we define the loss of first neighbors in the top monolayer as

$$\text{Weighted loss of first neighbors } (W) = \sum_{x=A,B} \left[ L_x \frac{D_{V,x} C_{S,x}}{\sum_{x=A,B} D_{V,x} C_{S,x}} \right]. \quad (3)$$

Here, $x$ is the atom type in the binary $AB$-system, $L_x$ is the fractional loss of the first neighbors of species $x$ at the surface and is defined as $(CN_V - CN_S)/CN_V$ with $CN$ being the coordination number and the subscript referring to either the bulk or the surface atom of type $x$, $D_V$ is the bulk diffusion coefficient, and $C_S$ is the fraction (or concentration) of atoms of a certain atom type on the surface. We note that $D_S/D_V$ is poorly correlated with $D_V$ at $T_g$ ($R^2 = 0.17$) and, hence, $D_V$ in Eq. 3 does not lead to any artificial correlation between $D_S/D_V$ and $W$. For most of the glasses in this study, $D_{V,A} C_{V,A} \gg D_{V,B} C_{V,B}$ (with $A$ and $B$ defined as the first and second elements listed in the MG compositions in Table 1) and $W$ is almost equal to $L_A$. Low values of $W$ qualitatively indicate that surface atoms have fewer missing bonds and vice versa. Fig. 4(b) shows that $W$ correlates to $D_S/D_V$ quite well over the whole range of MGs in this study. This behavior can be contrasted with silica and hydrogen-bonded organic glasses where the surface bonding maintains aspects of the bulk and therefore does not translate into dramatic surface enhanced diffusion.[7] For example, as discussed in Sec. I, for the mobile Si atoms in $SiO_2$, the number of Si-O bonds are very similar for surface and bulk atoms, leading to $D_S/D_V$ near 1.[12] In MGs, with all bonds between atoms being approximately equivalent in determining mobility, the fact that there are significant missing nearest-neighbor bonds at the surface leads to an increase of surface versus bulk mobility. We note that the correlation of $W$ with $D_S/D_V$ may be due to correlation of $W$ with both $D_{0,S}/D_{0,V}$ and $Q_V - Q_S$. It is easy to understand why $Q_V - Q_S$ might be correlated with $W$ as changing bond numbers will change binding strength which is expected to correlate with activation energies. Trends in $D_{0,S}/D_{0,V}$ are expected to correlate with anharmonicity of interatomic interactions since $D_0$ is related to anharmonicity,[31,34] and anharmonicity may correlate with bonding environment and therefore $W$, although this correlation is not immediately obvious. However, in Sec. S5 of the Supplementary Material, we show that $Q_V - Q_S$ and $D_{0,S}/D_{0,V}$ do not correlate with W independently, and it is only their combined contributions that correlate. This result is somewhat surprising and suggests that there is significant coupling between $D_{0,S}/D_{0,V}$ and $Q_V - Q_S$ in ways that require them to be integrated together to see the correlation with $W$.

### 4. Correlation of fragility to surface enhanced diffusion

Here, we aim to correlate $D_S/D_V$ with a glass property called fragility. To understand the fragility, we note that an important classification of glasses is in terms of the amount of slowdown when approaching $T_g$ during cooling.[35] This slowdown is generally quantified by tracking increasing viscosity (η) or decreasing diffusion coefficient. In experiments, the viscosity of supercooled liquids increases by more than ten orders of magnitude, and η = $10^{12}$ Pa s at $T_g$ by convention. *Strong* glass-formers like silica have an almost Arrhenius-like increase (or linear behavior in an Angell plot of log(η) versus $T_g/T$) in viscosity over a wide temperature range. MGs are relatively *fragile* and exhibit the viscosity rise in a narrow temperature range just below $T_g$. To characterize these differences, the fragility index (*m*) is introduced, which is defined as the slope of the Angell plot at $T_g$.[35] Strong and fragile glass-formers have low and high values of *m*, respectively. Fragility may be expected to be correlated with $D_S/D_V$ because from Fig. 1, $D_S ≈ D_V$ at the beginning of the supercooled state and the surface enhanced diffusion at $T_g$ is mostly dominated by the slowdown manifested by the bulk, characterized by the fragility. Another way to rationalize why fragility may underlie enhanced surface diffusion comes from the entropy theory of glasses that relates fragility to the frustration in atomic packing.[36–38] According to this theory, fragile systems have frustrated atomic packings. On the other hand, the surface is expected to have reduced packing frustration, thereby behaving like a strong glass. Accordingly, we can expect fragile systems to have large $D_S/D_V$ while enhanced surface diffusion in strong systems would be much milder. A strong correlation between fragility and $D_S/D_V$ could provide both insights into the surface diffusion physics and approaches to determine $D_S$ more efficiently using only bulk properties.

A measure of fragility can also be obtained by using the value of viscosity at $1.25T_g$,[7] with more fragile liquids having smaller values of η at $1.25T_g$. We will adopt this definition of fragility here, with the viscosity replaced by the structural relaxation time ($\tau$) of the system, as the latter is easier to calculate with MD. It has been shown using MD simulations that $\tau$ is proportional to η,[39] so the correlations obtained using $\tau$ should also occur when using η. As the relaxation times of glasses are different at their respective $T_g$ in our simulations, we will rescale $\tau$ by the value of $\tau$ at $T_g$ so that the rescaled value at $1.25T_g$ can be directly used as a measure of fragility. Fig. 5(a) shows the variation of scaled relaxation times in an Arrhenius plot for different MGs in this study, and Fig. 5(b) shows how $D_S/D_V$ correlates to our measure of fragility. A modest correlation can be observed, demonstrating that the surface enhanced diffusion increases with increasing fragility of the system. $Ni_{80}P_{20}$ is the main outlier in Fig. 5(b), and we expand on this below.

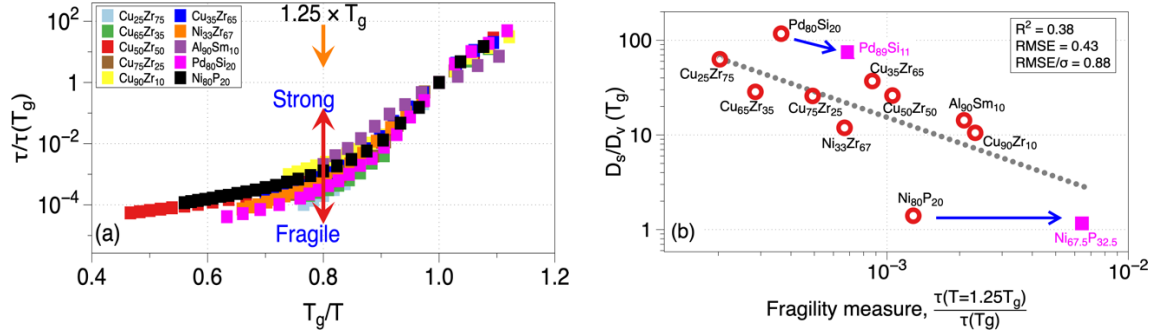

FIG. 5. (a) Variation of scaled relaxation time as a function of inverse temperature normalized to $T_g$ for different MGs in this study. The scaling of relaxation time is performed to be commensurate with identical viscosities at $T_g$ for glasses in experiments. (b) Plot of surface enhanced diffusion with respect to fragility of the glasses. The dotted line was fit to the original glass compositions, shown as open circles. Modified points that account for surface segregation as described in the text are shown as solid squares for $Pd_{80}Si_{20}$ and $Ni_{80}P_{20}$, the two glasses with the greatest surface segregation.

Fragility is a bulk property and does not capture the intricacies of the surface, which is different from the earlier metrics we have used to correlate to $D_S/D_V$. For example, the segregation of an element to the surface, which is a consequence of missing bonds intrinsic to the surface and could impact the overall $D_S$, likely cannot be predicted by fragility. Among the compositions studied in this work, $Pd_{80}Si_{20}$ shows the highest value of $D_S/D_V$ and the concentration of the mobile element, palladium, increases from 80% in the bulk to ~89% at the surface (~11% increase). $Ni_{80}P_{20}$ has the lowest $D_S/D_V$ in our simulations and, in this case, the concentration of the immobile element, phosphorous, increases from 20% to ~32.5% at the surface (~63% increase). These changes are illustrated in the inset of Fig. 6(a). Segregation thus might be playing a role in both these cases in either enhancing or diminishing the surface diffusion, an effect that the bulk fragility cannot capture. To quantify the effect of segregation, we recognize that segregation plays no role when either (i) the concentration of elements is identical in the bulk and the surface, or (ii) the mobilities of the two elements are equal at the surface. Based on these 2 criteria, we quantify the effect of segregation on surface diffusion as the product of two terms as

$$\text{Segregation measure } (S) = \sum_{x=A,B}\left[(C_{S,x} - C_{V,x})\left(\frac{D_{S,x} - D_{S,avg}}{D_{S,avg}}\right)\right]. \qquad (4)$$

Here, $C_V$ is the fraction (or concentration) of atoms of a certain atom type in the bulk, $D_{S,avg}$ is the average surface diffusion of the two elements. Both terms inside the square brackets for $S$ in Eq. 4 vary in the range from -1 to 1. A positive value for $S$ implies that the mobile element is segregated to the surface, thereby enhancing $D_S$ and vice versa. Fig. 6(a) shows the variation of $S$ for different MGs studied in this work, with $Ni_{80}P_{20}$ exhibiting the most negative values, and therefore likely the largest segregation-induced reduction of $D_S$. To assess the hypothesis that segregation may be impacting the correlation with fragility, we consider the correlation between the residuals of a linear fit of $D_S/D_V$ to fragility (deviation from the dotted line shown in Fig. 5(b)) and our segregation measure ($S$ from Eq. (4) and shown in Figure 6(a)). The residuals vs. $S$ in Figure 6(b) show a modest but definite correlation, suggesting that the deviations of $D_S/D_V$ from a linear function of fragility are caused to a significant extent by surface segregation.

The change in composition of the surface could modify $D_V$, the fragility, or both. To assess these changes, we have simulated using periodic boundary conditions bulk glasses with the surface compositions for the glasses with the highest and lowest values of $S$ in Fig. 6(a). For $Pd_{80}Si_{20}$, the surface composition is $Pd_{89}Si_{11}$, and for $Ni_{80}P_{20}$ the surface composition is $Ni_{67.5}P_{32.5}$. $D_V$ and fragility are evaluated for the bulk glass with the surface composition, while $D_S$ is evaluated from the slab with the original composition. Both the diffusivities are measured at $T_g$ of the original composition. $D_S/D_V$ vs. fragility evaluated this way, and shown as solid squares in Fig. 5(b), is much closer to the trend line for the surface composition of $Ni_{80}P_{20}$ while no improvement is observed for $Pd_{80}Si_{20}$. On the other hand, both $D_S$ and $D_V$ can also be evaluated at $T_g$ of the segregated surface composition and in this case, we observed (not shown) the surface composition of $Pd_{80}Si_{20}$ agreeing very well with the trend line in Fig. 5(b) while $Ni_{80}P_{20}$ shows no improvement. We conclude that fragility can be a useful indicator to estimate surface enhanced diffusion, but that surface segregation can lead to a departure from the estimated behavior.

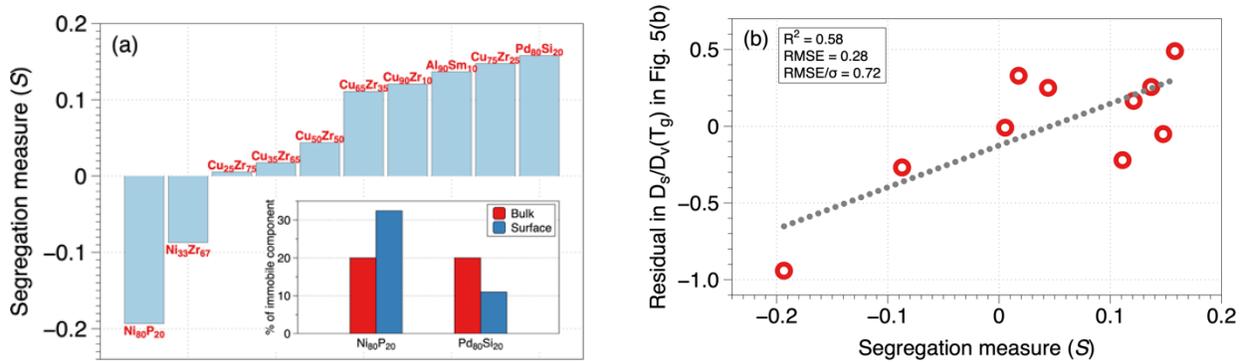

FIG. 6 (a) Quantifying the presence of segregation at the surface for the MGs involved in this study. (inset) Comparison of the concentration of immobile components of the bulk and surface in $Pd_{80}Si_{20}$ (immobile component Si) and $Ni_{80}P_{20}$ (immobile component P). (b) Correlation between $S$, defined in Eq. 4, and the deviation in Fig. 5(b) between the calculated $D_S/D_V$ values and those obtained from a best fit to fragility.

In Sec. S4 of the Supplementary Material, we illustrate a method of estimating *experimental-like* $D_S/D_V$ from MD diffusion values. An important result from the experimental bulk and surface diffusion coefficients shown in Ref.[7] (Fig. 5 therein) is that there is a linear relation between $\log D_S$ and $\log D_V$. Here, we assume that this linear trend between $\log D_S$ and $\log D_V$ also applies to MGs based on the observation that there is an excellent fit observed between these parameters for most of our MGs in the supercooled and glassy state. Also, for the organic glass-formers polystyrene and *o*-terphenyl (OTP), it was recently shown that experimental and molecular dynamics (MD) bulk and surface diffusion coefficients show a linear relationship between $\log D_S$ and $\log D_V$.[40] We estimate the experimental surface enhanced diffusion $D_S/D_V$ by extrapolation from the linear trend between $\log D_S$ and $\log D_V$ in Fig. S4 of the Supplementary Material.

## IV. CONCLUSIONS

We have simulated multiple binary metallic glasses (MGs) using molecular dynamics with the goal to identify different factors that influence enhanced surface diffusion $D_S/D_V$ at $T_g$. While $D_V$ exhibits a rapid slowdown of dynamics when approaching $T_g$ during cooling, only a moderate slowdown of surface dynamics is observed. $D_S/D_V$ at $T_g$ varies by two orders of magnitude for the compositions studied. Both the surface energy and the fraction of missing bonds for surface atoms show a significant correlation with the enhanced surface diffusion, implying that the surface bond breaking directly translates into higher mobility, unlike in network- or hydrogen-bonded organic glasses. Furthermore, segregation of elements at the surface is observed in some cases and can significantly impact the surface enhanced diffusion. Fragility, a measure of the slowdown dynamics in the bulk close to $T_g$, also correlates with $D_S/D_V$, although this correlation appears to fail for some cases with high segregation. We applied extrapolation methods to approximate the surface enhanced diffusion expected for typical experimentally cooled glasses and found that the extrapolated $\ln(D_0)$ vs. $Q$ values for the surface follows the previously established trend for bulk glasses.

## SUPPLEMENTARY MATERIAL

See the supplementary material for the time evolution of MSD of atoms in the glassy state in $Cu_{50}Zr_{50}$; decay of intermediate scattering function to estimate relaxation time ($\tau$); evaluation of the decoupling parameter in the fractional Stokes-Einstein equation; estimating experimental-like surface enhanced diffusion $D_S/D_V$ from MD simulations; and correlation of activation energy change for diffusion with $W$.

## DATA AVAILABILITY

The data for all the figures in the Manuscript and Supplementary Information are shared as digital files in the Supplementary Information and on Figshare at 10.6084/m9.figshare.13682542. For any other relevant data, contact AA at vkannama@ncsu.edu.

## ACKNOWLEDGMENTS

The authors are grateful to the Extreme Science and Engineering Discovery Environment (XSEDE), which is supported by National Science Foundation grant number OCI-1053575, and the Center for High Throughput Computing at UW-Madison for the computing resources. This work was supported by the University of Wisconsin Materials Research Science and Engineering Center (DMR-1720415).

# Supporting Material

## Factors correlating to enhanced surface diffusion in metallic glasses

Ajay Annamareddy,[1] Yuhui Li,[2] Lian Yu,[2,3] Paul M. Voyles,[1] and Dane Morgan[1]

[1]Department of Materials Science and Engineering, [2]School of Pharmacy and [3]Department of Chemistry, University of Wisconsin-Madison, Madison, WI 53706, USA.

### S1. Time evolution of mean-squared displacement (MSD) of atoms in bulk $Cu_{50}Zr_{50}$

Figure S1 shows the variation of mean-squared displacement (MSD) of atoms with time in bulk $Cu_{50}Zr_{50}$ at temperatures 700 K (= $T_g$) and 640 K (lowest bulk temperature studied in $Cu_{50}Zr_{50}$). The relaxation time ($\tau$) of $Cu_{50}Zr_{50}$ at these temperatures are also indicated in the figure.

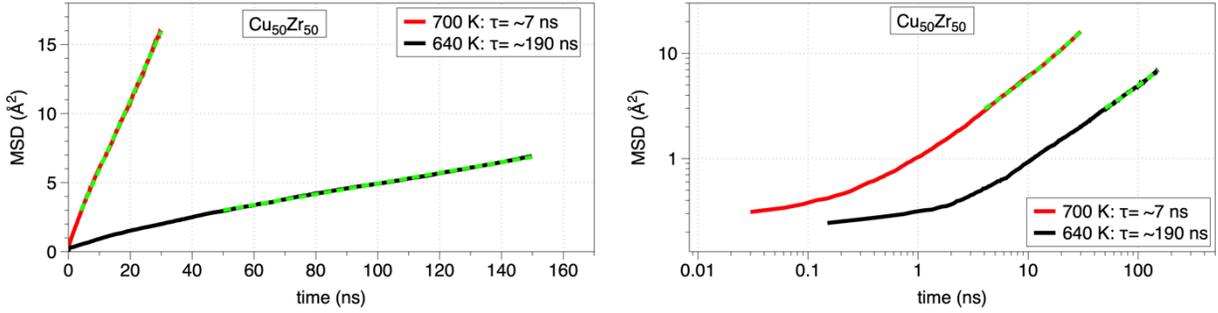

FIG. S1. Time evolution of MSD in $Cu_{50}Zr_{50}$ at 700 K and 640 K in (left) linear-linear and (right) log-log plots. The dotted lines indicate the best linear fit at longer times. The relaxation times of the system ($\tau$) at the two temperatures are also indicated in the legend.

### S2. Decay of Intermediate Scattering Function $F_S$ in bulk $Cu_{50}Zr_{50}$

Figure S2 shows the time variation of $F_S$ with time in bulk $Cu_{50}Zr_{50}$ at 700 K. The dotted line represents the well-known Kohlrausch–Williams–Watts (KWW) function, $f(t)$ = A exp[-$(t/\tau)^\gamma$], with A and $\gamma$ as the parameters. For temperatures when the decay of $F_S$ is less than 1/e, the KWW fit is used to identify the relaxation time, $\tau$.

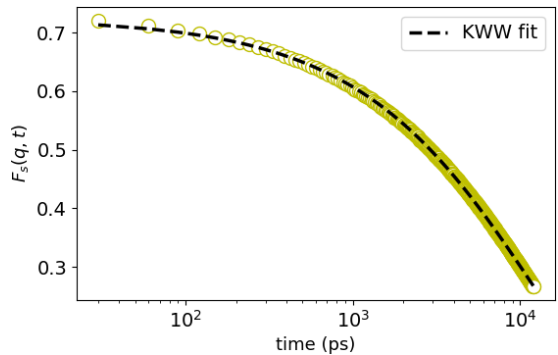

FIG. S2. Time evolution of the intermediate scattering function, $F_S$ in $Cu_{50}Zr_{50}$ at 700 K (= $T_g$).

### S3. Estimation of the decoupling parameter ($\zeta$) in the fractional Stokes-Einstein equation

Figure S3 shows $D/T$ plotted against $1/\tau$ for $Cu_{50}Zr_{50}$ at 700 K and the slope of the linear fitting shown (red solid line) determines the decoupling parameter, $\zeta$.

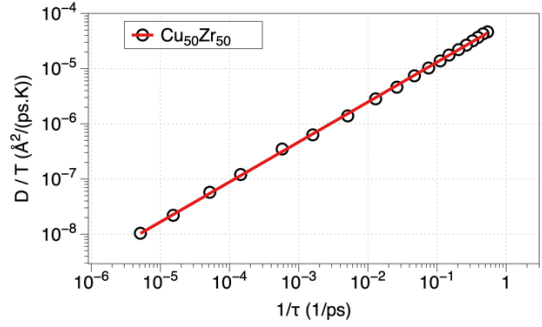

FIG. S3. Variation of $D/T$ versus inverse relaxation time in $Cu_{50}Zr_{50}$ at 700 K (= $T_g$). The slope of the linear fit gives the decoupling parameter ($\zeta$) of the fractional Stokes-Einstein equation.

### S4. Estimating experimental $D_S/D_V$ from MD simulations

An important result from the experimental bulk and surface diffusion coefficients shown in Ref.[1] (Fig. 5 therein) is that there is a linear relation between log $D_S$ and log $D_V$. Here, we assume that this linear trend between log $D_S$ and log $D_V$ also applies to MGs based on the observation that there is an excellent fit observed between these parameters for most of our MGs in the supercooled and glassy state. The $R^2$ values for the linear fit in different MGs are shown in Table S1. Fig. S4 shows log $D_S$ against log $D_V$ for $Cu_{50}Zr_{50}$.

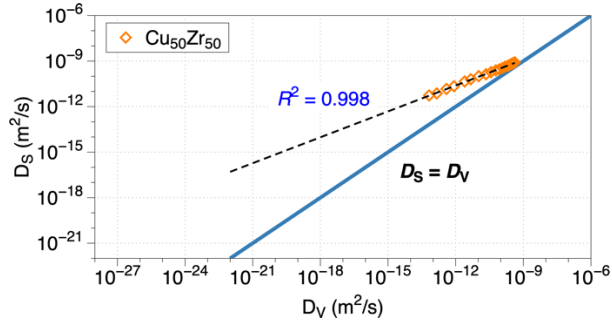

FIG. S4. Plot of MD evaluated $D_S$ against $D_V$ for $Cu_{50}Zr_{50}$. The dotted line shows the linear fit between log $D_S$ and log $D_V$. The dotted line is extended to extrapolate the $D_S$ corresponding to a typical experimental $D_V = 10^{-22}$ m²/s.

For the organic glass-formers polystyrene and $o$-terphenyl (OTP), it was recently shown that experimental and molecular dynamics (MD) bulk and surface diffusion coefficients show a linear relationship between log $D_S$ and log $D_V$.[2] Based on this, we now estimate the experimental surface enhanced diffusion $D_S/D_V$ from the linear trend between log $D_S$ and log $D_V$ in Fig. S4 by extrapolation. Note that we do not perform the extrapolation if the linear fit is not good enough, and we have chosen a cutoff of $R^2 > 0.98$ for when we perform the extrapolation (only $Ni_{80}P_{20}$ fails this criterion). We assume a fixed $D_V$ value at $T_g$ of $10^{-22}$ m²/s, which is consistent with typical experimental values. Fig. S4 shows the fit of log $D_S$ against log $D_V$ for $Cu_{50}Zr_{50}$ extrapolated to $D_V = 10^{-22}$ m²/s. The surface diffusion is enhanced over the bulk by ~$10^5$ after extrapolation.

The enhanced surface diffusion for all the alloys in this work determined the same way are given in Table S1. The surface enhancement factor at the laboratory $T_g$ obtained by extrapolation is in broad agreement with the limited experimental results.[3,4] The extrapolated experiment bulk and surface diffusion values at experimental $T_g$ (Table 1 in the paper gives the values of $T_g$) were used to estimate a more realistic $D_0$ for MD activation energies, which are shown as solid symbols in Fig. S5. The new $(Q, \ln(D_0))$ pairs more closely follow the experimental observations compared to the earlier estimates derived solely from MD simulations. Especially, the similar $\ln(D_0)$ vs. $Q$ trends in extrapolated surface MD results, extrapolated bulk MD results, and experimental results suggests that the procedure followed in Fig. S4 provides a realistic estimate of $D_S$ and that the linear relationship of $\ln(D_0)$ vs. $Q$ found for bulk diffusion also approximately holds for surface diffusion in metallic glasses.

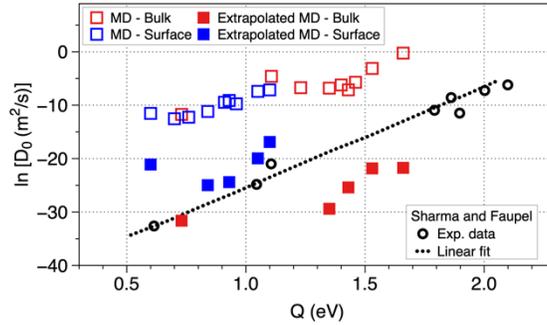

FIG. S5. The pre-exponentials corresponding to the filled squares are obtained by using the experimental $T_g$, a typical experimental $D_V$ of $10^{-22}$ m²/s for all glasses, while the surface diffusion-related $D_0$ are obtained from the extrapolation scheme discussed above.

Table S1. Details of the lowest temperature investigated in this work for bulk dynamics, length of the simulation, and the corresponding relaxation time at this temperature for different metallic glass compositions. The fitting scheme illustrated in FIG. S4 is used to obtain the extrapolated $D_S/D_V$ values (for $D_V = 10^{-22}$ m²/s) shown. The $R^2$ value of the fit is also shown.

| Composition ($A_XB_{100-X}$) | Lowest temperature studied | Simulation time | Relaxation time ($\tau$) | Extrapolated $D_S/D_V$ | $R^2$ value |
|---|---|---|---|---|---|
| $Cu_{25}Zr_{75}$ | 660 K | 250 ns | 495 ns | 4.8e5 | 0.995 |
| $Cu_{35}Zr_{65}$ | 640 K | 200 ns | 297 ns | 2.7e5 | 0.995 |
| $Cu_{50}Zr_{50}$ | 640 K | 150 ns | 195 ns | 5.0e5 | 0.998 |
| $Cu_{65}Zr_{35}$ | 680 K | 150 ns | 134 ns | 8.6e5 | 0.996 |
| $Cu_{75}Zr_{25}$ | 680 K | 180 ns | 135 ns | 1.3e6 | 0.994 |
| $Cu_{90}Zr_{10}$ | 660 K | 150 ns | 45 ns | 2.5e4 | 0.998 |
| $Pd_{80}Si_{20}$ | 680 K | 400 ns | 1350 ns | 5.1e6 | 0.989 |
| $Ni_{80}P_{20}$ | 520 K | 250 ns | 616 ns | - | 0.976 |
| $Al_{90}Sm_{10}$ | 580 K | 30 ns | 13 ns | 1.0e5 | 0.987 |

**S5. Correlation of activation energy change for diffusion ($Q_V$ - $Q_S$) with weighted loss of first neighbors $W$**

To verify if the change of activation energy for surface diffusion ($Q_V - Q_S$) and the ratio of pre-factors $D_{0,S}/D_{0,V}$ are correlated to the weighted loss of first neighbors $W$ (which is related to coordination number change and given by Eq. 3 in the main manuscript), we plot in Fig. S6 the correlation between these quantities. It can be observed that $W$ does not correlate well with both ($Q_V - Q_S$) and $D_{0,S}/D_{0,V}$.

FIG. S6. Plot of (a) activation energy change between bulk and surface diffusion versus $W$, and (b) ratio of pre-exponential factors $D_{0,S}/D_{0,V}$ versus $W$, across several MGs in our study illustrating almost zero correlation ($R^2 \sim 0$) between them.